# NOISE ANALYSIS FOR TWO QUANTUM CRYPTOGRAPHY PROTOCOLS

Sindhu Chitikela

**Abstract.** This paper presents noise analysis for the two-stage and the three-stage quantum cryptographic protocols based on random polarization rotations. The noise model used is that of uniform distribution of error over a certain small range that is associated with each link without regard for the source of the error. The noise in different links is taken to be independent. Advantages of the use of these protocols in low intensity laser systems are given.

1. **Introduction**

In quantum cryptography information is transmitted in the form of photons or light particles. These information carriers obey laws of quantum physics. A photon which is a quantum state exhibits the property that in general it is a superposition of mutually exclusive attributes and the unknown polarization of a photon cannot be cloned.

Quantum cryptography is a secure way to distribute a random secret key between two parties. Once the key is shared through a quantum channel, the information is then transferred through a public channel using classical cryptographic techniques. Quantum cryptography, which is generally implemented as in the BB84 protocol [1], [2], makes it possible to implement a one-time pad under certain conditions. Quantum states are susceptible to interaction with the environment as well as errors in the state generation process and in the orientation of the basis states.

Here we limit our attention to the effect of noise on the two-stage and the three-stage protocols for key distribution or encryption of data [3]. The next section describes how the BB84 protocol operates in the presence of noise. The next sections describe the two-stage and the three-stage quantum cryptography protocols and present noise analysis under certain assumptions. A comparison between the multi-stage protocols and BB84 is also provided.

2. **Quantum Cryptography in the Presence of Noise**

Consider first the question of information in a photon. Represented as a qubit $(a|0\rangle + b|1\rangle)$, the photon will collapse to $|0\rangle$ or $|1\rangle$ but since this collapse is random, it would not communicate any useful information to the receiver. The maximum information will be communicated to the receiver if the sender prepares the photon in one of the two orthogonal states $|0\rangle$ or $|1\rangle$, which the receiver will be able to determine upon observation. In the BB84 protocol, the photon is polarized using either rectilinear bases (horizontal and vertical bases) or diagonal bases ($-45^0$ and $+45^0$ bases) to represent qubits. A photon generated represents a qubit after it is passed through a linear polarizer. A horizontally polarized photon or a photon that is polarized by $-45^0$ represents a qubit 1. Similarly a vertically polarized photon or a photon that is polarized by $+45^0$ represents a qubit 0.

In the BB84 protocol, at the sender's site (Alice), a thin beam of light which emits single photon at a time is sent through linear polarizer to generate polarized photons that represent a bit 0 or 1. Alice codes a quantum bit using either a rectilinear base state or a diagonal base state. The base states are chosen randomly. The resultant stream comprises of horizontally, vertically polarized photons and diagonally polarized photons each representing a qubit of the information that Alice wants to send to Bob. In this case of key distribution, the information sent is a pseudorandom sequence that is used to compose a key.

On receiving the stream of polarized photons, Bob measures the property of the photon by using two interchangeable polarizing beam splitters where one of them allows Bob to distinguish between the horizontal and vertical polarization, the other allows him to distinguish between $-45^0$ and $+45^0$ polarization. He then uses photon detectors to know the arrival of the photon. The choice of polarizing beam splitters is randomly made. If Bob uses a polarizing beamsplitter compatible with the polarization choice of Alice, he can read the bit as either 0 or 1 based on the polarization. If Bob uses a polarizing beamsplitter incompatible with the polarization choice of Alice, he cannot get any information about the state of polarization. As stated earlier, here the quantum state is a superposition of two mutually exclusive properties. There is a 50 % probability of it being a bit that represents 0 or 1.

Once Alice is done with sending the information, Bob announces the sequence of polarizing beamsplitters that he used. Alice then compares this sequence with the sequences of bases that she used and tells Bob to note the beamsplitters that he used correctly. Finally those bits that are obtained from the correctly selected beamsplitters are considered and this forms the final key called the sifted key.

In the BB84 protocol, the effect of noise is combated by first estimating the noise rate on the public channel and then extracting the reconciled key from the raw key. First, Alice and Bob apply an agreed upon random permutation to their respective raw keys. The raw key is broken into blocks of length *x*, where the value of x is chosen so that it is most likely that the block contains no more than one error. For each of these blocks, and for other sub-blocks, Alice and Bob publicly compare parity checks, in a process so that erroneous bits are located and deleted. Each time parities are compared, an agreed upon bit is deleted from the chosen key sample. If the parity should not agree, a binary search strategy is used to locate and delete the error.

The correctness of the raw key can also be communicated between Alice and Bob by the use of cryptographically strong hash functions of their respective raw keys. The ideas of obtaining reconciled key starting from a raw key can be used for non-BB84 protocols also.

## 3. Kak's two-stage protocol

The two-stage protocol is described in passing in his paper on the three-stage protocol [3]. At an abstract level, two-stage protocol can be explained in a simple scenario of exchange of a bag of money between Alice and Bob. Imagine a situation where Alice sends an amount of money to Bob, say X dollars. Bob either adds or subtracts some amount to this, say (X+Y) dollars and sends this to Alice. Alice now subtracts her amount from the bag and knows the value added by Bob. If the money in the bag is not counted in passage in both directions, then the protocol is secure.

This situation can be related to the two-stage protocol as a scenario where Alice sends a linearly polarized photon that is randomly rotated through an angle $\theta$ to Bob. It is Bob who has to decide on the key to be shared and he either rotates the photon further by $90^0$ or by $0^0$ that is no rotation. In the third step, Alice rotates the photon by $-\theta$ and knows the rotation performed by Bob and hence knows the data chosen by Bob same as Alice knows the amount added by Bob. So, by knowing the rotation made by Bob, Alice knows the binary value decided by Bob to form a key. Hence the key negotiation can be done in just two stages.

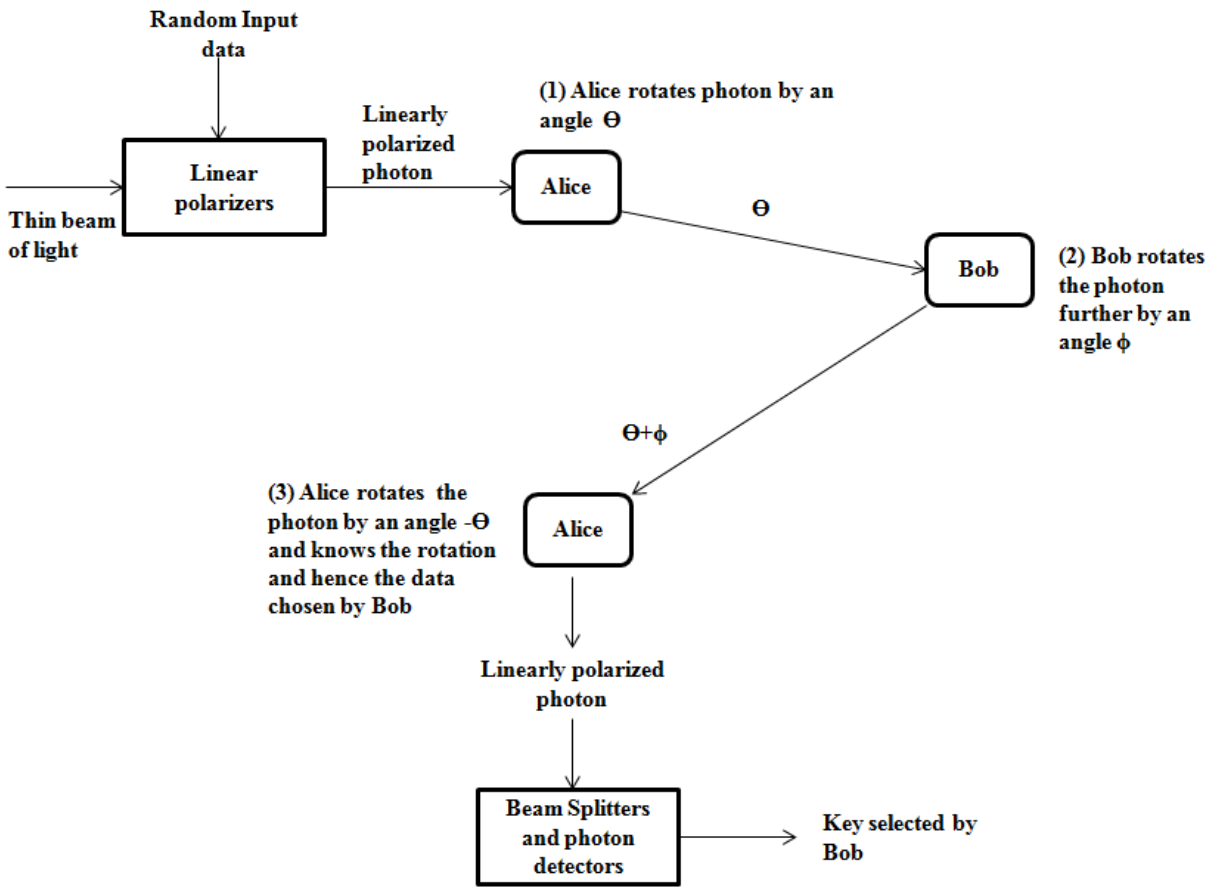

Figure 1 Implementation of the two-stage protocol

Once, the entire key bits are known by Alice, Bob now computes the hash value of the key and sends it to Alice. Similarly Alice computes the hash value of the key received and sends it to Bob. They compare the hash values received with the one they have computed and verify that the correct key is shared. This step ensures the secure distribution of secret keys. The working of the two-stage protocol is shown in Figure 1.

4. **Kak's three-stage protocol**

Unlike BB84 protocol, the entire communication between Alice and Bob remains quantum at each stage that is both the key distribution and the further information transfer takes place in a quantum channel. The working of Kak's three-phase protocol is explained in the following section.

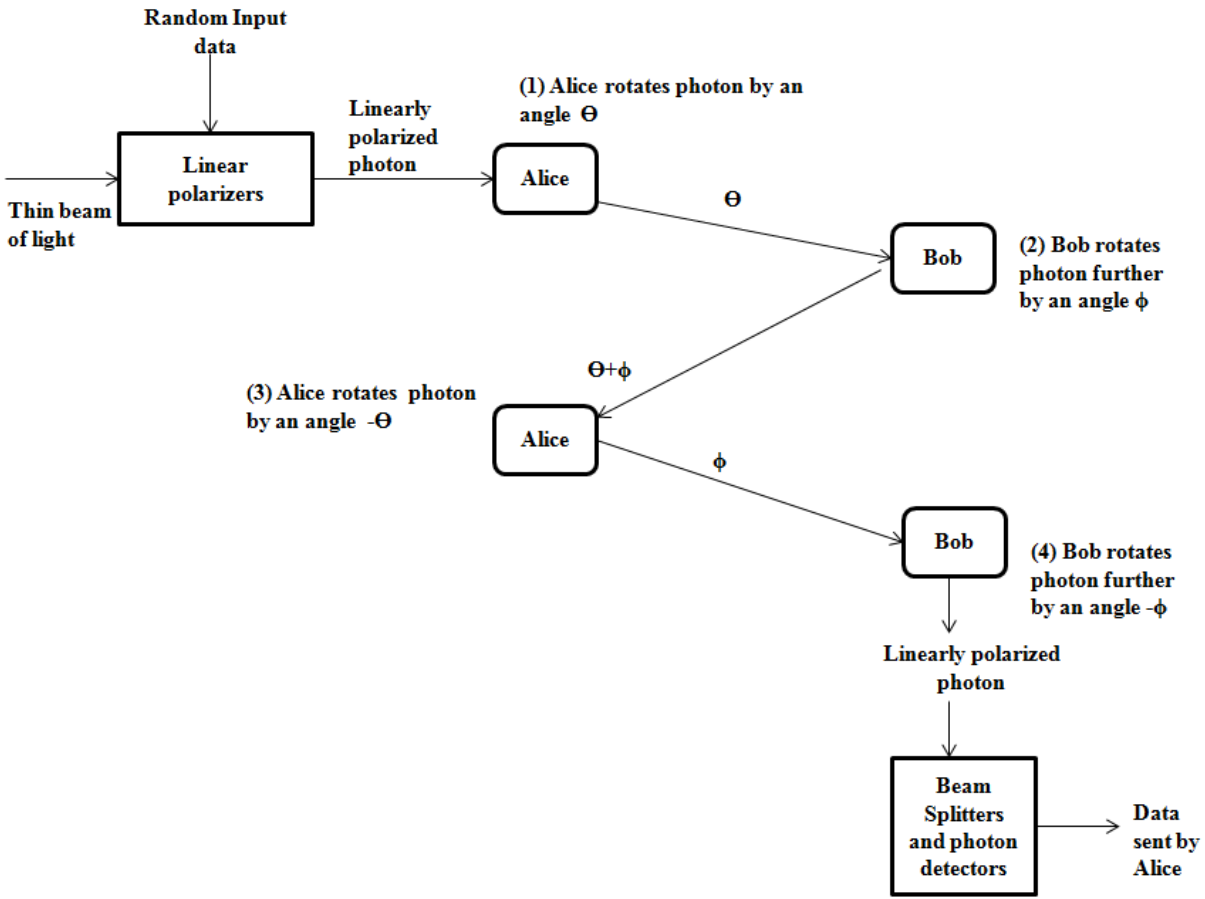

Figure 2 Implementation of Kak's three-stage protocol

A simple scenario of transfer of a box from Alice to Bob explains the three-stage protocol at a simple level. Consider a situation in which Alice puts her own lock on a box and sends it to Bob. Bob then puts his own lock on the box and sends it to Alice. Now, Alice removes her lock from the box and sends it to Bob. Bob now removes his own lock from the box and finally he can open and see what is in the box. The same scenario is the three-stage protocol where the locks are nothing but the secret transformations that each party does.

Kak's three-stage protocol involves secret rotation transformations on the photons. Alice codes a quantum bit by applying a secret transformation of some random angle on a polarized photon. In the initial stage, Alice rotates the polarized photon, X through a secret, random angle θ and sends it to Bob. In the second stage, Bob further rotates this photon through an angle of ϕ. In the final stage, Alice inverses the transformation by re-rotating this photon by the same angle θ and sends it to Bob. Bob inverses the transformation that he applied on the photon by re-rotating by angle, ϕ. Finally Bob receives the photon X that Alice intended to send. Figure 2 shows the working of the three-stage protocol.

### 5. Noise Analysis of Two-Stage Protocol and Three-Stage Protocol

In both the two-stage and three-stage protocols, there is a possibility that the equipment that rotates the photon through certain angle has error in it. The error can be at Alice's site or at Bob's site or it may be at both the sites. The error can also be in the quantum channel through which the photon is transferred. We will lump the error from various sources to a uniform probability distribution function in each communication link. The probability density function of error at Alice's or Bob's site is uniform with the measure of angle $x$ ranging from -0.1 to +0.1 radians and is represented in the Figure 3.

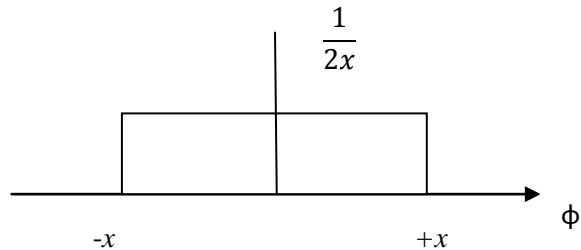

Figure 3 Probability Density Function of Error

Therefore, the probability of error is obtained by the following expression where ϕ ranges from −$x$ to +$x$ radians. We have

$$\int_{-x}^{+x} \frac{1}{2x} \sin^2\phi \, d\phi = \frac{1}{2} - \frac{1}{4x}\sin 2x$$

which is approximately equal to $\frac{1}{3}x^2 - \frac{1}{15}x^4$ where $x$ ranges from -0.1 to +0.1 . The graph obtained by taking different values is shown in the Figure 4. The y-axis represents the probability of error and the x-axis represents the value of $x$.

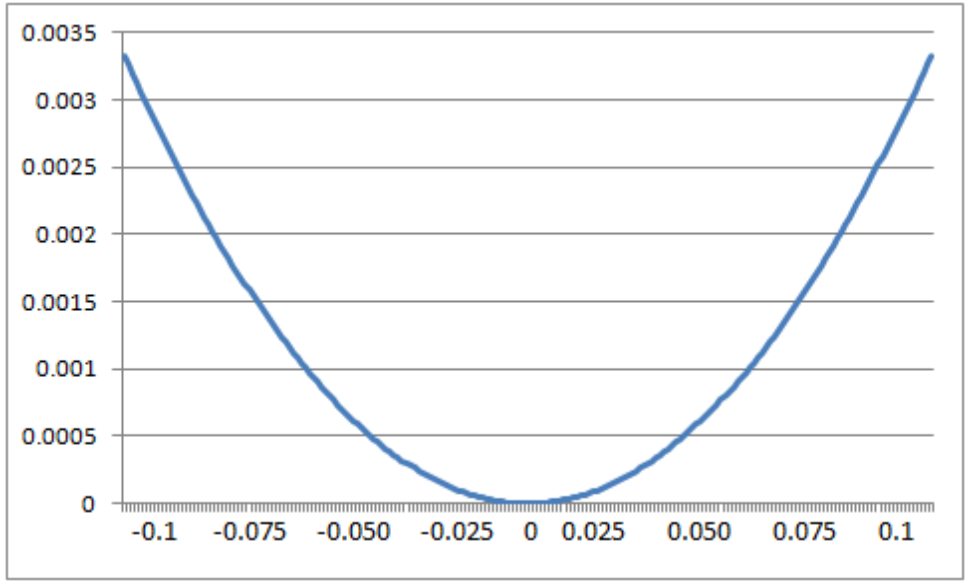

Figure 4 Probability of error based on the value of *x* in radians

When the error exists at both the sites, then the cumulative effective of noise is as shown in the Figure 5.

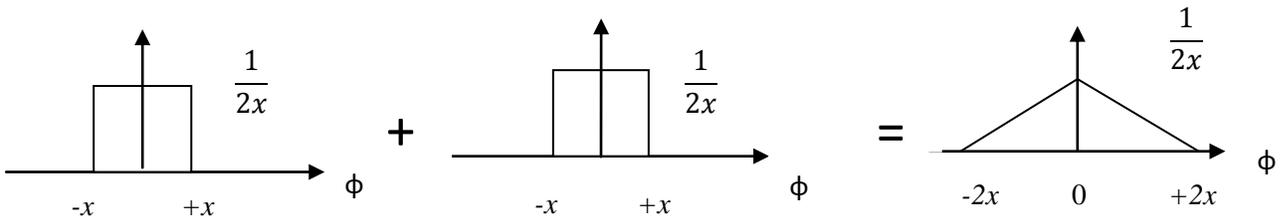

Figure 5 Cumulative effective of noise at both the sites

Therefore, the probability of error is obtained by the following expression where ϕ ranges from −*x* to +*x* radians.

$$2 \int_0^{2x} \left(-\frac{1}{4x^2}\phi + \frac{1}{2x}\right)(sin\phi)^2 d\phi = \frac{2x^2}{3}$$

where *x* ranges from -0.1 to +0.1 . The graph obtained by taking different values is shown in the Figure 6. The y-axis represents the probability of error and the x-axis represents the value of x.

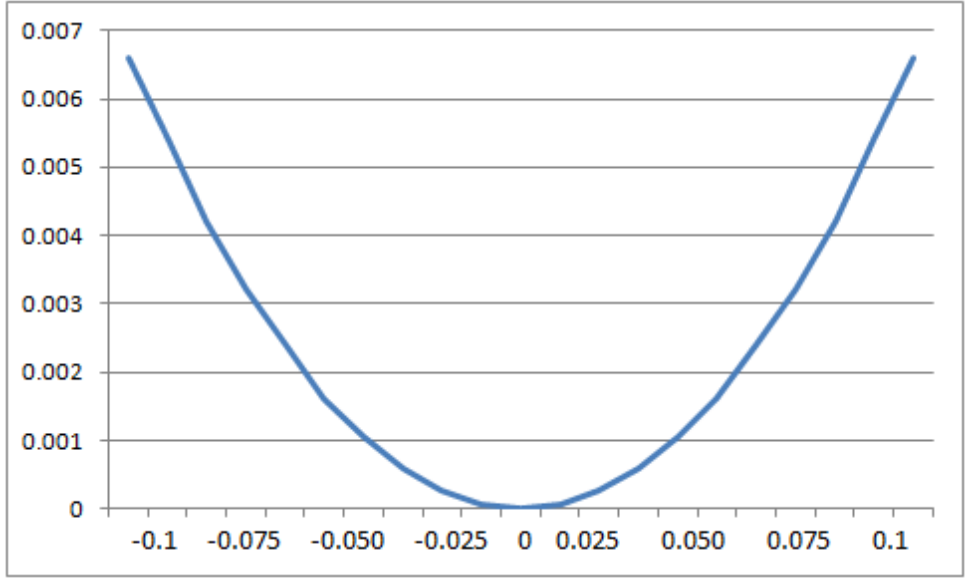

Figure 6 Probability of error based on the value of *x* in radians

We notice that the probability of error in two stages is roughly equal to twice of that in one stage. Likewise, the probability of error in three stages is roughly three times the error in a single stage.

Given this error rate, one can use standard methods of key purification to obtain the reconciled key from the raw copy.

### 6. Comparison of BB84 protocol and the Kak protocols

Apart from the faked-state attack [6],[7], the main weakness of BB84 protocol is that single photons are not easy to produce, and the duplicate photons can be used by the eavesdropper to reconstruct the key. Hence, the attacker can siphon off the photons when they are transferred between Alice and Bob. Moreover, as the photons are siphoned off only at one step, the intensity of the output at the receiver's end is not affected. There is also the problem in generating single photons [8] as well as having single photon detectors.

This is not the case with Kak's multistage protocols. In order to know the angles $\theta$ and $\phi$, Eve has to siphon off the photons in all the stages which can result in significant decrease in the intensity of the output. Hence, the receiver can easily identify the attack. Nevertheless, practical implementation of this system creates its own difficulties [9]-[12]. The security of single-photon rotation system has recently been presented [13]. A modification of the three-stage protocol to catch active eavesdroppers was recently presented [14].

Although it is generally claimed that quantum key distribution is unconditionally secure, Yuen has recently argued against that position [15].

## 7. Conclusions

This paper presents noise analysis for the two-stage and three-stage protocols of quantum cryptography, as well as the problem with attacks due to difficulty of single photon sources and detectors for the BB84 protocol. The noise model used is that of uniform distribution of error over a certain small range that is associated with each link without regard for the source of the error. The noise in different links is taken to be independent. Advantages of the use of these protocols in low intensity laser systems are given.

**Acknowledgement.** This research was supported in part by research grant #1117068 from the National Science Foundation.